\documentclass{article}
\usepackage[english]{babel}
\usepackage{authblk}

\usepackage[letterpaper,top=2cm,bottom=2cm,left=3cm,right=3cm,marginparwidth=1.75cm]{geometry}

\usepackage[sort&compress,round,semicolon,authoryear]{natbib}
\usepackage{hyperref}
\hypersetup
{
 unicode=false,     
 pdftoolbar=true,    
 pdfmenubar=true,    
 pdffitwindow=false,   
 pdfstartview={FitH},  
 pdftitle={Urban Density and Equity of Access to Social Services},  
 pdfauthor={Kerry A. Nice, Mark Stevenson},   
 pdfsubject={Subject},  
 pdfcreator={Creator},  
 pdfproducer={Producer}, 
 pdfkeywords={keywords}, 
 pdfnewwindow=true,   
 colorlinks=true,    
 linkcolor=red,     
 citecolor=blue,    
 filecolor=magenta,   
 urlcolor=cyan      
}

\usepackage{subcaption}
\usepackage{array}
\usepackage{pdfpages}
\newcolumntype{H}{@{}>{\lrbox0}l<{\endlrbox}}

\usepackage{orcidlink}

\title{Urban Density and Equity of Access to Social Services in Australian Urban Areas}
\author[1]{Kerry A. Nice\orcidlink{0000-0001-6102-1292}}
\author[1]{Mark Stevenson\orcidlink{0000-0003-3166-5876}}
\affil[1]{Transport, Health, and Urban Systems Research Lab, Faculty of Architecture, Building, and Planning, University of Melbourne, VIC, Australia}
\date{\today}

\begin{document}
\maketitle

\begin{abstract} 

To measure access to social services (primary health care, early childhood care/education, and public transport), we created two social service access indexes (SSPT and SSI) for Australian capital cities. We show that only two cities, Melbourne and Sydney, have some limited characteristics of a compact or 15-minute city, but only in the city centres and inner city areas where population densities are highest and have less low density housing types. In the outer suburban and peri-urban areas, as well as across all of the remaining cities, proximity to social services is poor and residents suffer the consequences of spatial inequity.





\end{abstract}


\section{Questions}

~~~~As Australian cities undergo growth in population and hence, urbanisation, the challenges to deliver access to services posed by historic planning decisions has become increasingly difficult; from early settlement urban form emphasising sprawling green-field self-sufficiency, Australian urban areas sprawled outward from highly centralised capital cities (in both form and administrative and commercial function) \citep{wilkinson_federalism_2022} following radial public transportation links and preferencing suburban home ownership and low density \citep{troy_structure_2004}. While some policy efforts in the 1990s sought to revitalise urban centres, they met with mixed success, with some increases of inner city population densities, but often resulting in density hot spots clustered around transport links \citep{coffee_visualising_2016}. The results are uneven distribution of access to services with considerable inequalities bourne by residents in peri-urban areas \citep{nice_how_2024,stevenson_urban_2025}. Consequences of these spatial inequities can include social inequalities and poverty \citep{harpham_urban_1997}, increased health risks (i.e. obesity observed to be 2.3x higher in outer-urban and regional areas \citep{australian_institute_of_health_and_welfare_australias_2018}), and disadvantage in access to services such as education, health or transport \citep{dikec_justice_2001}.

There has been growing interest recently in the benefits of compact cities \citep{breheny_sustainable_1992,burton_compact_2000,burton_measuring_2002,daneshpour_compact_2011,elkin_reviving_1991,newman_cities_1989}, often defined as urban areas with high densities of population and infrastructure that facilitate short journeys to amenities (schools, retail, and workplaces) through mixed land use \citep{stevenson_urban_2016}. Related, are 15- or 20-minute cities, designed so that essential services and amenities are within a 15 minute active transport (cycling or walking) journey from local areas \citep{bruno_universal_2024}.

We aim to test access to essential services across Australian capital cities and create social service indexes (SSPT and SSI). We ask the question, can any Australian city be considered or exhibit characteristics of a compact or 15-minute city? Additionally, what can the spatial distributions of these indexes show about the links between urban form, infrastructure and population density, demographics, and the distribution of spatial advantage/disadvantage across Australian capital cities?

\section{Methods}

We combine three main datasets that provide locations of primary health services, childcare centres, and proximity and frequencies of public transportation services. For additional context, we link to these demographic data (housing types, population density, socioeconomic status) from the \cite{australian_bureau_of_statistics_2021_2021} 2021 Community Profiles and Socio-Economic Indexes for Areas (SEIFA) \citep{australian_bureau_of_statistics_socio-economic_2021}.

Health providers data was obtained from the National Health Services Directory, maintained by Health Direct Australia\footnote{https://www.healthdirect.gov.au/}, and includes providers, service types (i.e. pharmacy, family practice, mental health, etc.), opening hours, billing practices, service address, and latitude/longitude. Childcare facility data was obtained from the Australian Children's Education \& Care Quality Authority (ACECQA) National Register\footnote{https://www.acecqa.gov.au/resources/national-registers} and includes providers, provider street address, number of places, hours of operation, and service quality ratings. Data clean up for both involved geolocating missing latitude/longitude locations from street addresses. The number of public transport services (bus, train, and tram) per week at each latitude/longitude transport stop were calculated using General Transit Feed Specification (GTFS)\footnote{https://developers.google.com/transit/gtfs/reference} files for all Australian public transport providers \citep{mobilitydata_io_mobility_2024}. GTFS files are generated by public transport agencies to provide timetables and geographic locations of routes and stops.

The number of services in the primary health and childcare categories and public transport access (frequency and proximity of services) within a flat earth distance of 800m (a distance approximating a 15 to 20-minute walk on the street network) was calculated from 100m grids of points across each of the urban areas. Census and SEIFA data was attached to each point from the Australian Statistical Geography Standard (ASGS)\citep{abs_digital_2021} SA1 (statistical areas designed to include a population of 200-800 persons) in which the point was located.

Two indexes were created, SSPT and SSI. To generate the SSPT index, counts of access to public transport (proximity and frequency), general practitioners and pharmacies and early childhood education and care within 800 metres were normalised and scaled 0 to 1 and all indicators were summed across each SA1. The SSI index only included the primary health and childcare services. 

The resulting indexes could range from 0 (no access to services) to very high accessibility, 3 or 2 for SSPT and SSI respectively. Finally, we aggregate the index results along with demographic characteristics to SA4 levels to show broader trends across the capital cities. SA4s are the largest sub-State statistical regions and are designed to contain at least 100,000 persons and best represent the boundaries of regions that contain both where residents reside and work \citep{abs_digital_2021}.

\section{Findings}
\begin{figure*}[ht!]
\centering
\begin{subfigure}[t]{0.3\textwidth}
\includegraphics[trim={480 0 483 20},clip,scale=0.11]{"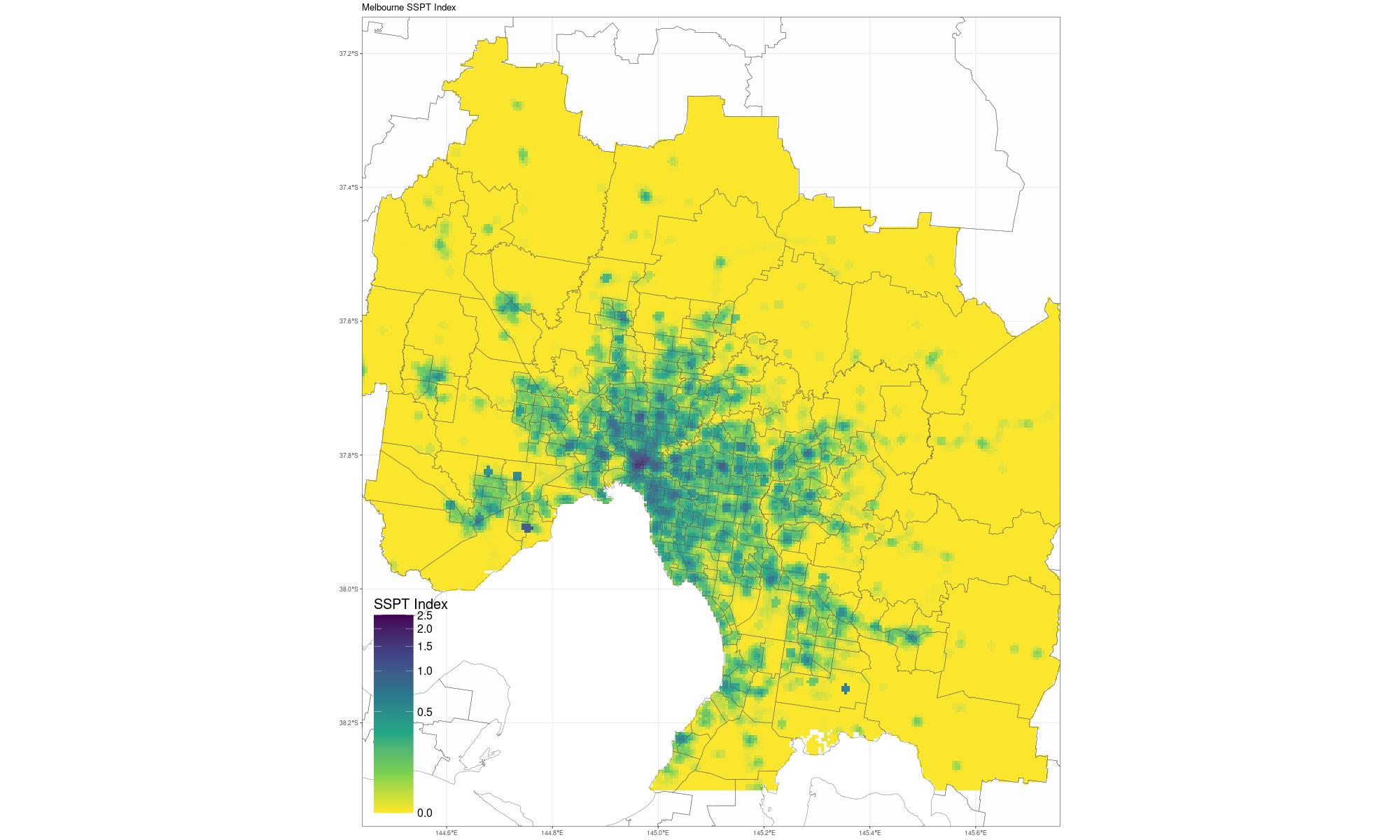"}\caption{Melbourne SSPT}
\end{subfigure} %
\hfill
\begin{subfigure}[t]{0.3\textwidth}
\centering
\includegraphics[trim={447 0 455 20},clip,scale=0.11]{"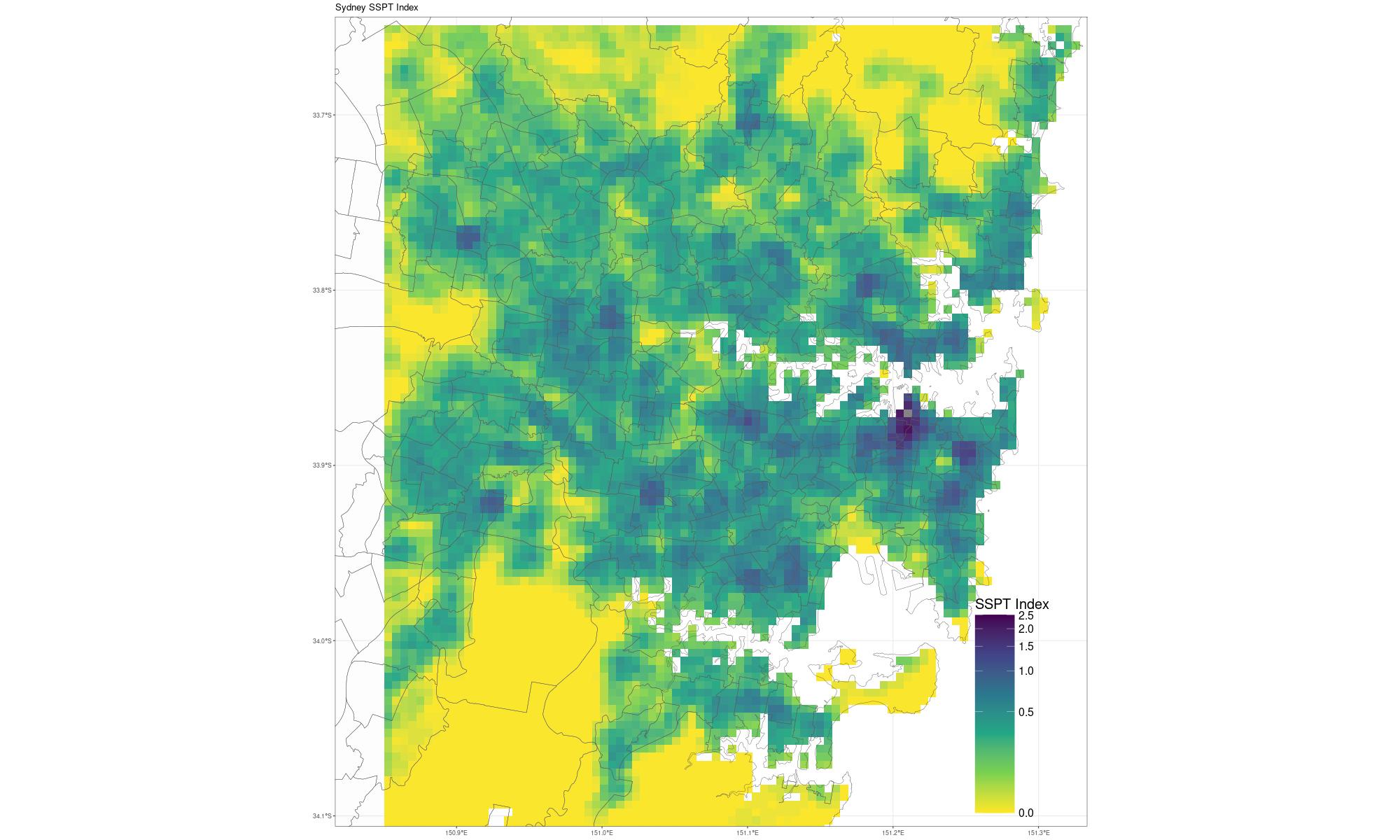"} \caption{Sydney SSPT}
\end{subfigure}
\hfill
\begin{subfigure}[t]{0.3\textwidth}
\centering
\includegraphics[trim={470 0 470 20},clip,scale=0.11]{"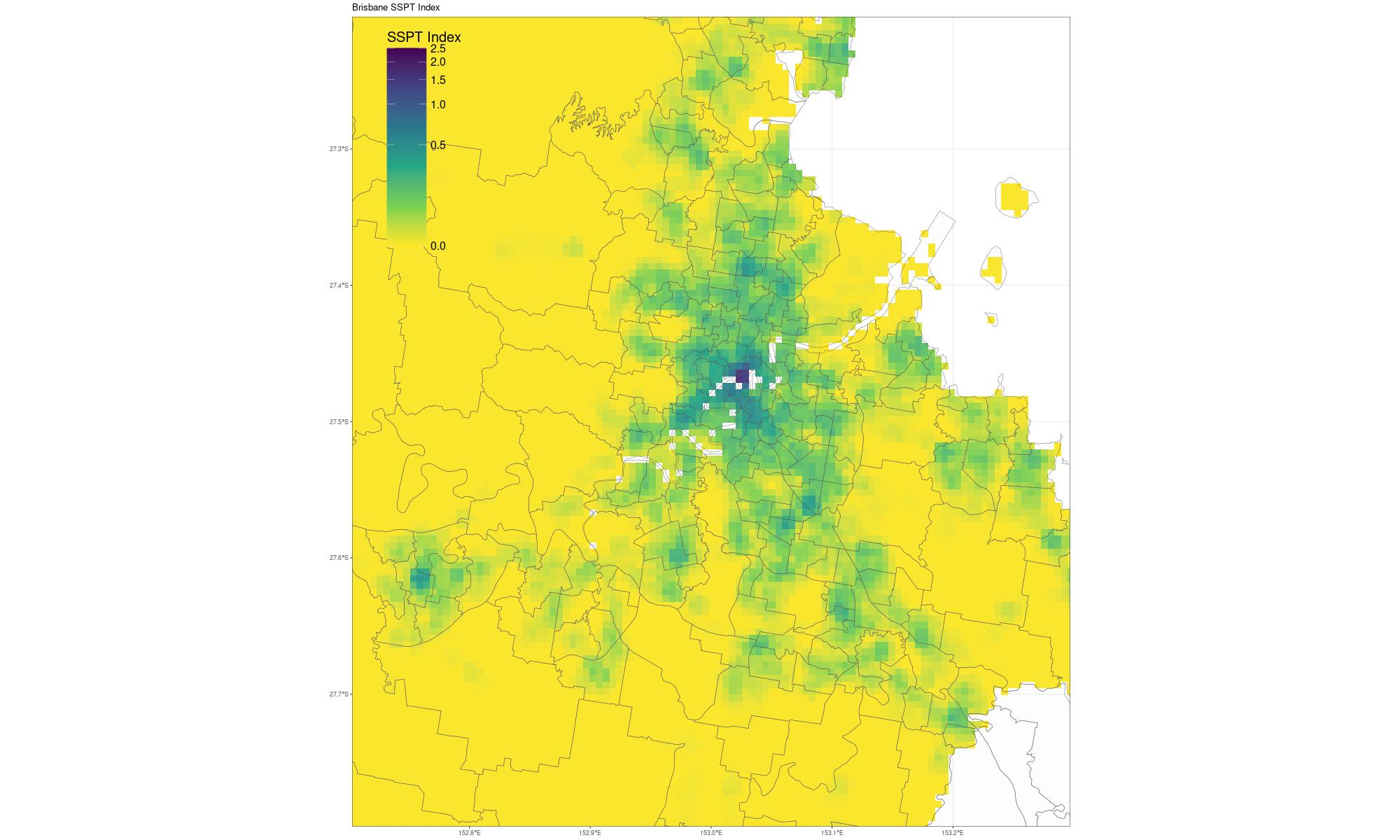"}  \caption{Brisbane SSPT}
\end{subfigure}
\centering
\begin{subfigure}[t]{0.3\textwidth}
\includegraphics[trim={480 0 483 20},clip,scale=0.11]{"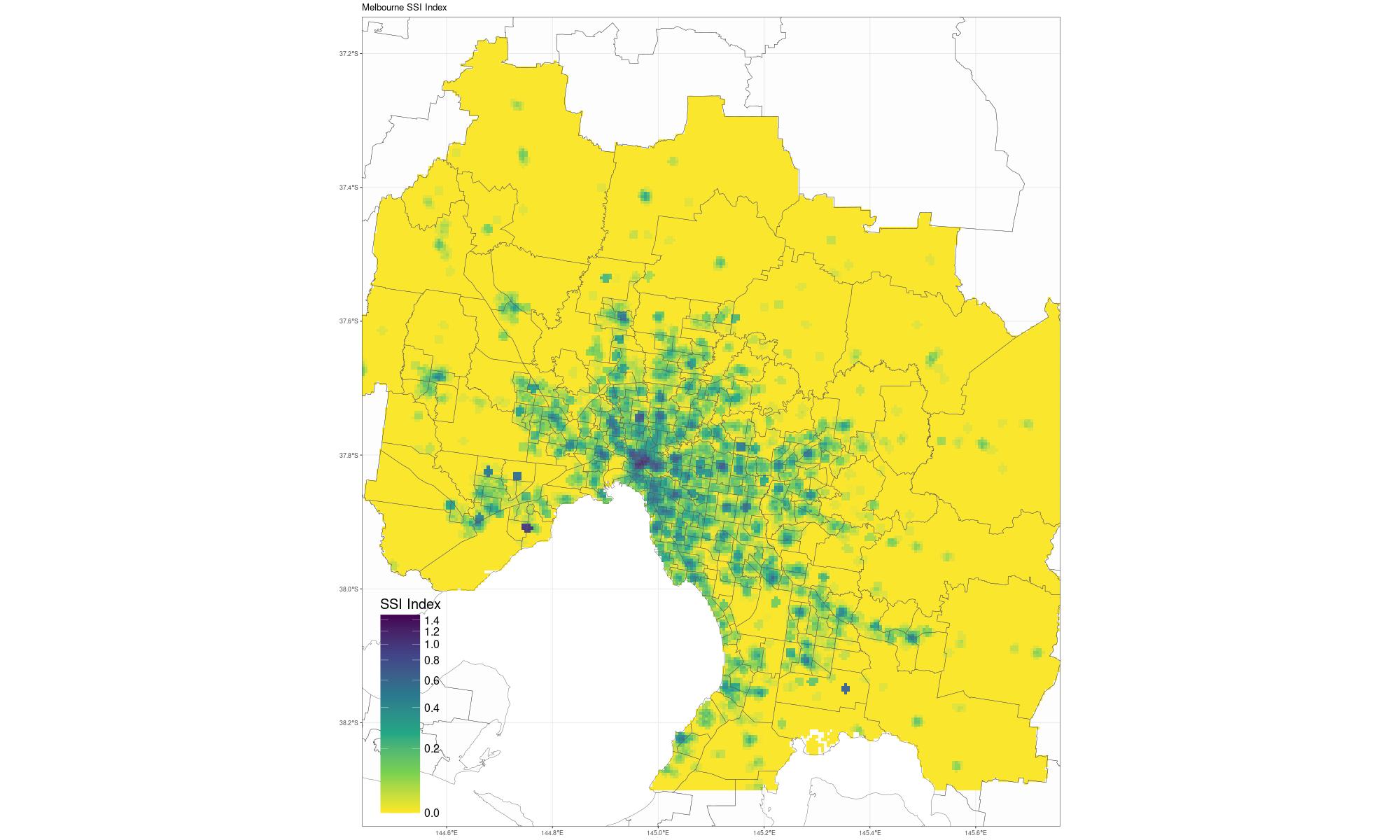"}\caption{Melbourne SSI}
\end{subfigure} %
\hfill
\begin{subfigure}[t]{0.3\textwidth}
\centering
\includegraphics[trim={447 0 455 20},clip,scale=0.11]{"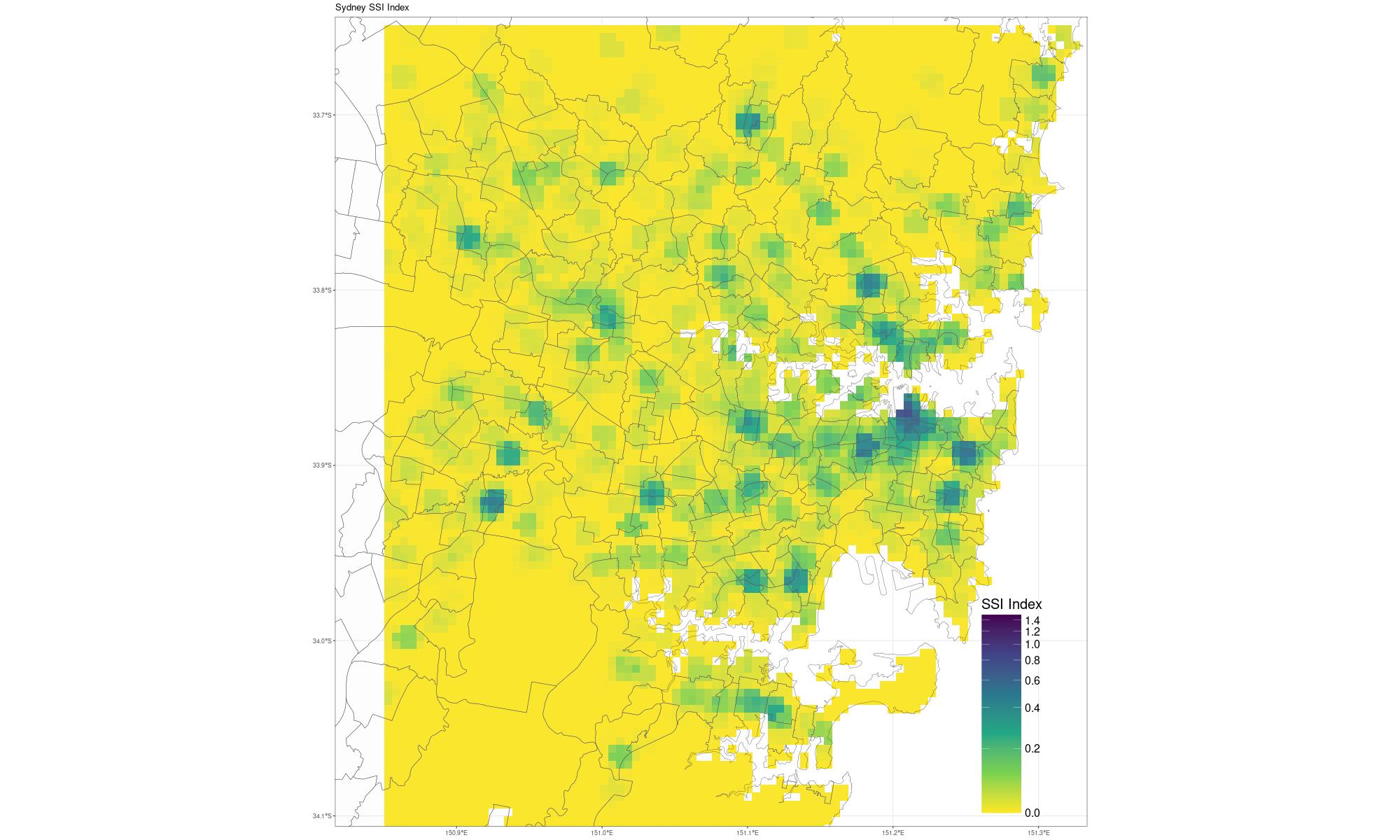"} \caption{Sydney SSI}
\end{subfigure}
\hfill
\begin{subfigure}[t]{0.3\textwidth}
\centering
\includegraphics[trim={470 0 470 20},clip,scale=0.11]{"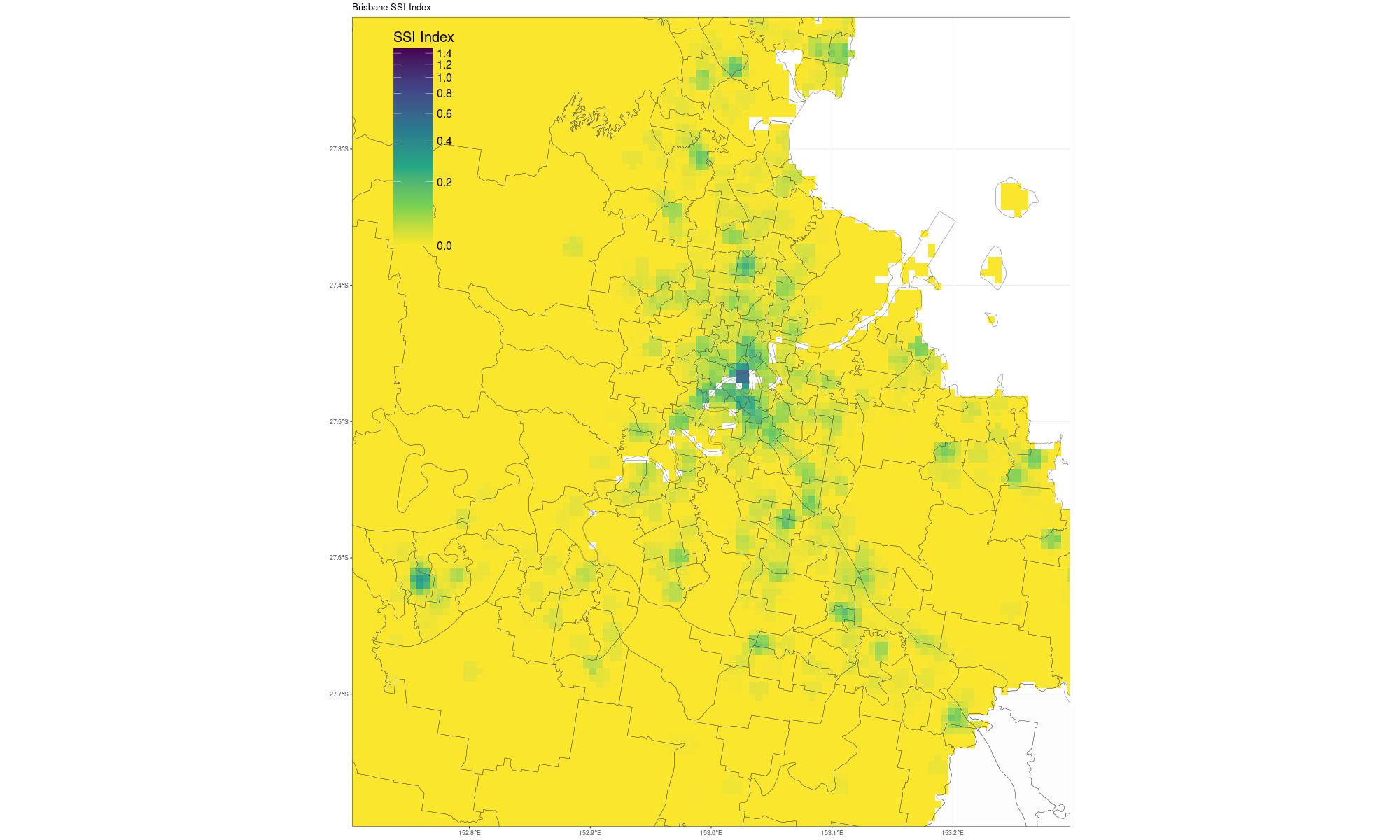"}  \caption{Brisbane SSI}
\end{subfigure}
\centering
\begin{subfigure}[t]{0.3\textwidth}
\includegraphics[trim={430 0 490 20},clip,scale=0.11]{"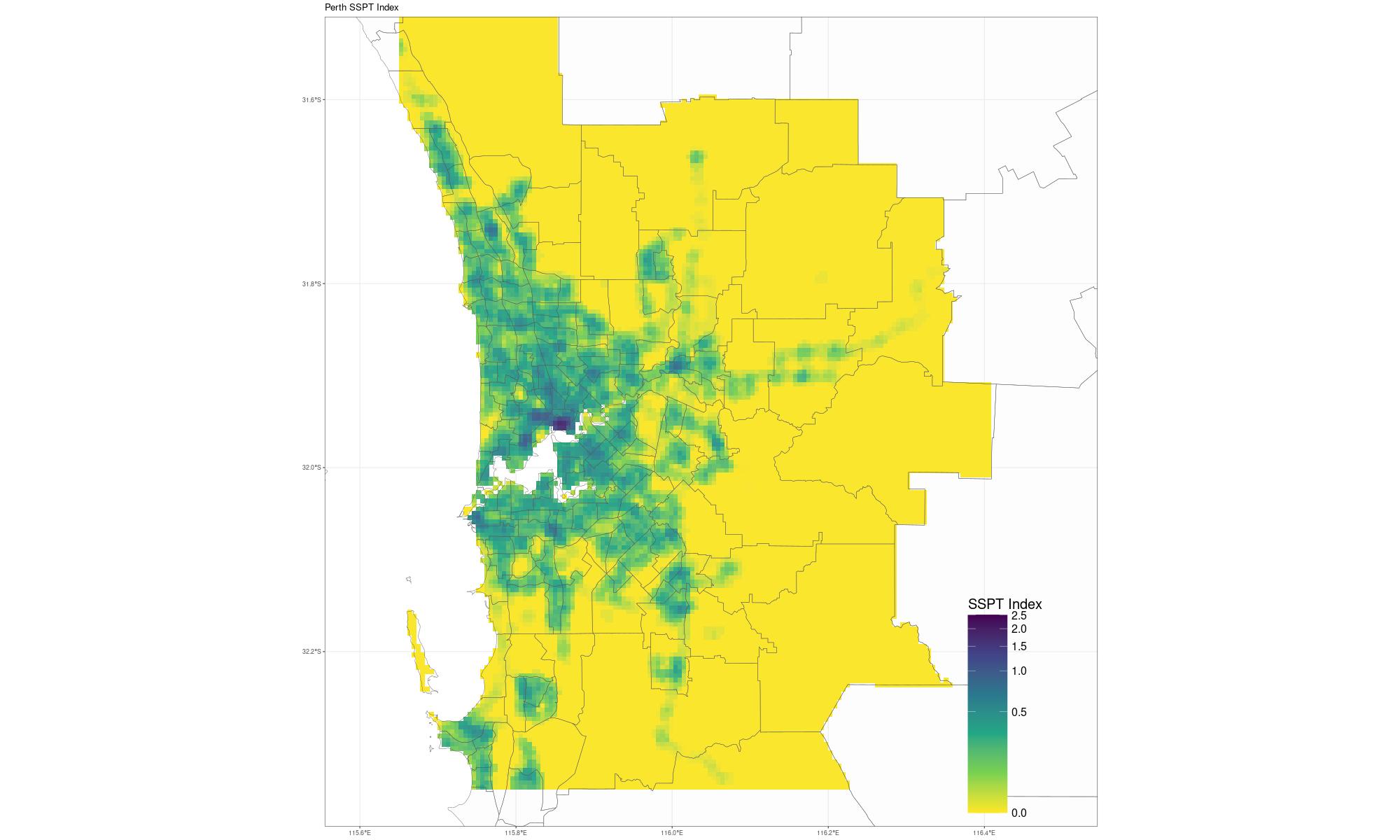"} \caption{Perth SSPT}
\end{subfigure} %
\hfill
\begin{subfigure}[t]{0.3\textwidth}
\centering
\includegraphics[trim={260 0 260 20},clip,scale=0.11]{"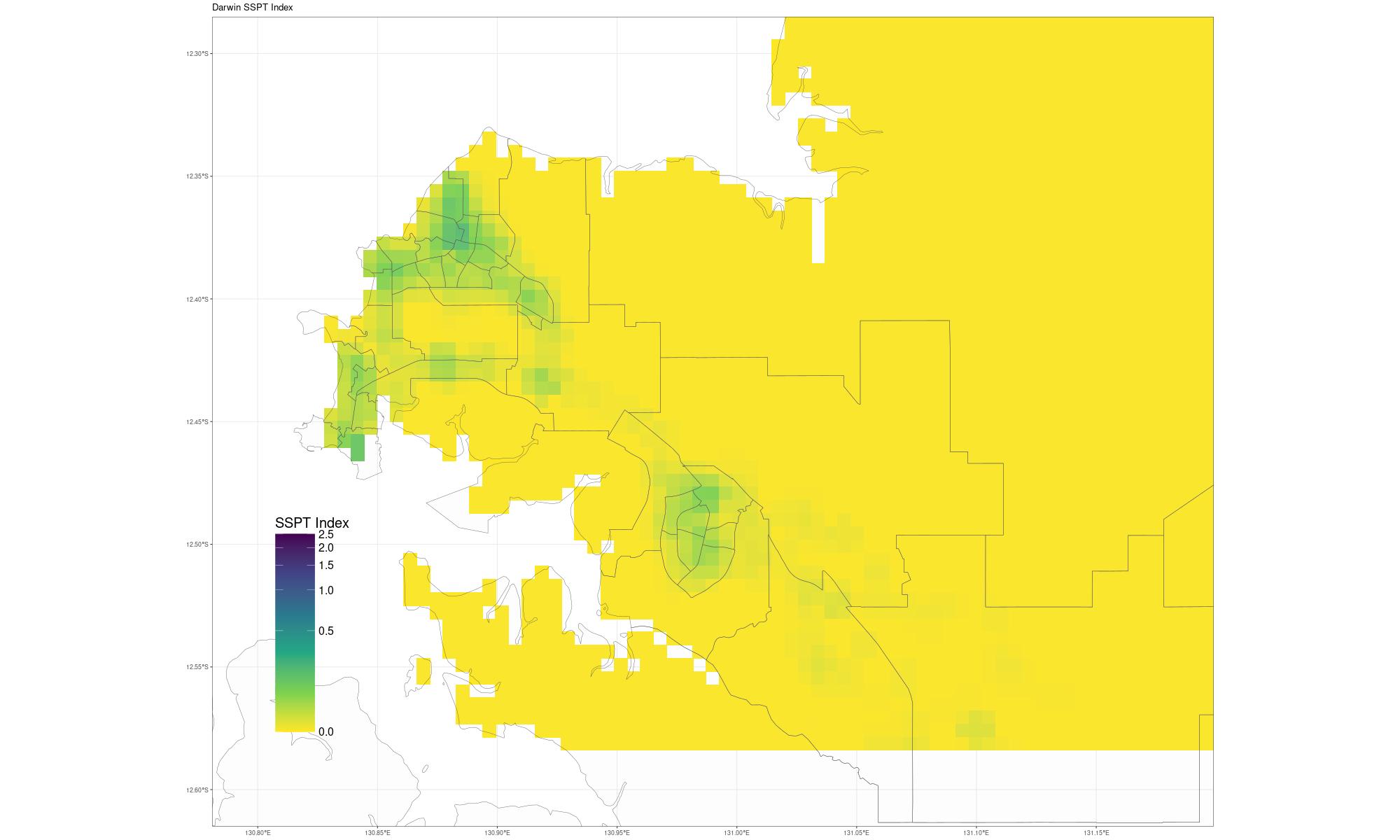"} \caption{Darwin SSPT}
\end{subfigure}
\hfill
\begin{subfigure}[t]{0.3\textwidth}
\centering
\includegraphics[trim={630 0 630 20},clip,scale=0.11]{"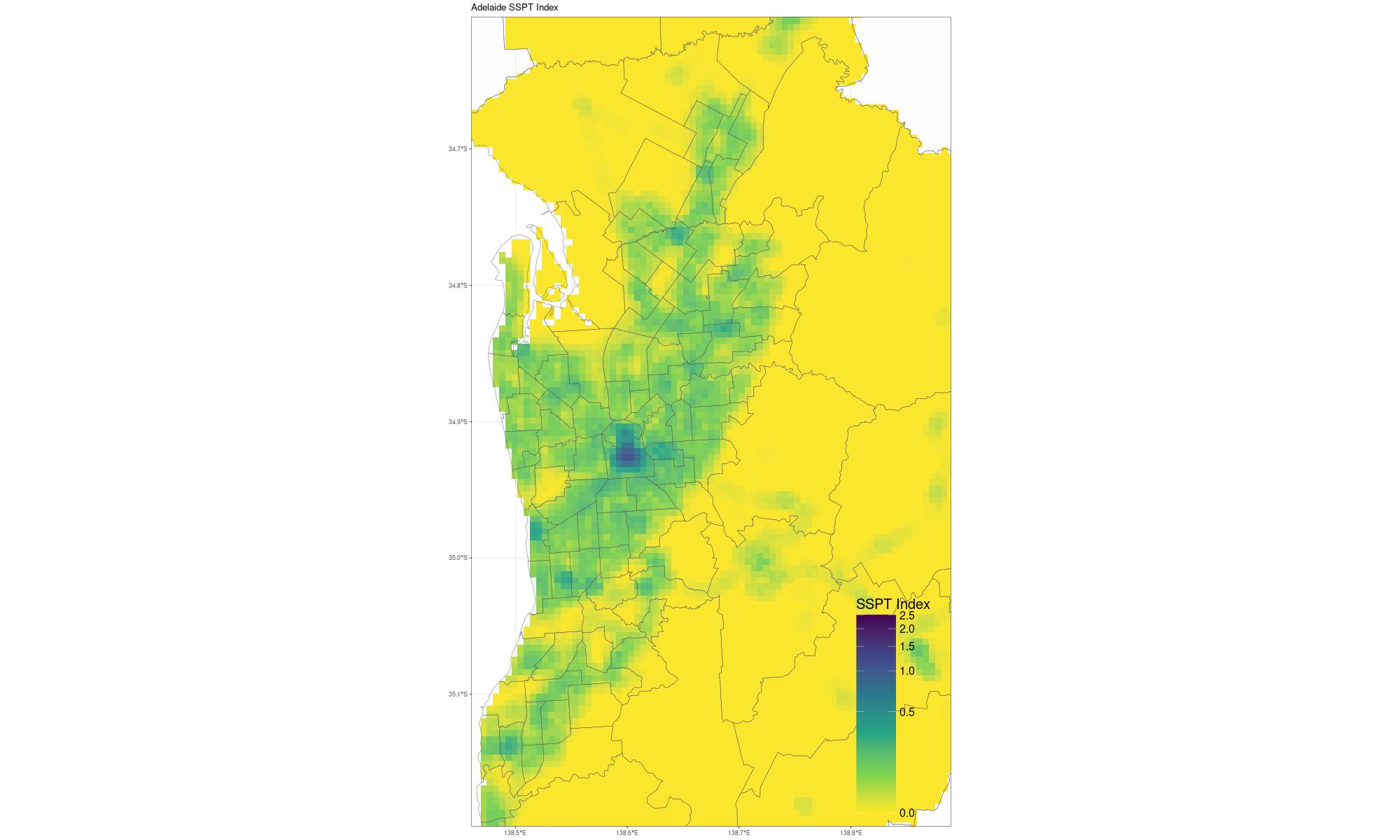"}\caption{Adelaide SSPT}
\end{subfigure}
\centering
\begin{subfigure}[t]{0.3\textwidth}
\includegraphics[trim={430 0 490 20},clip,scale=0.11]{"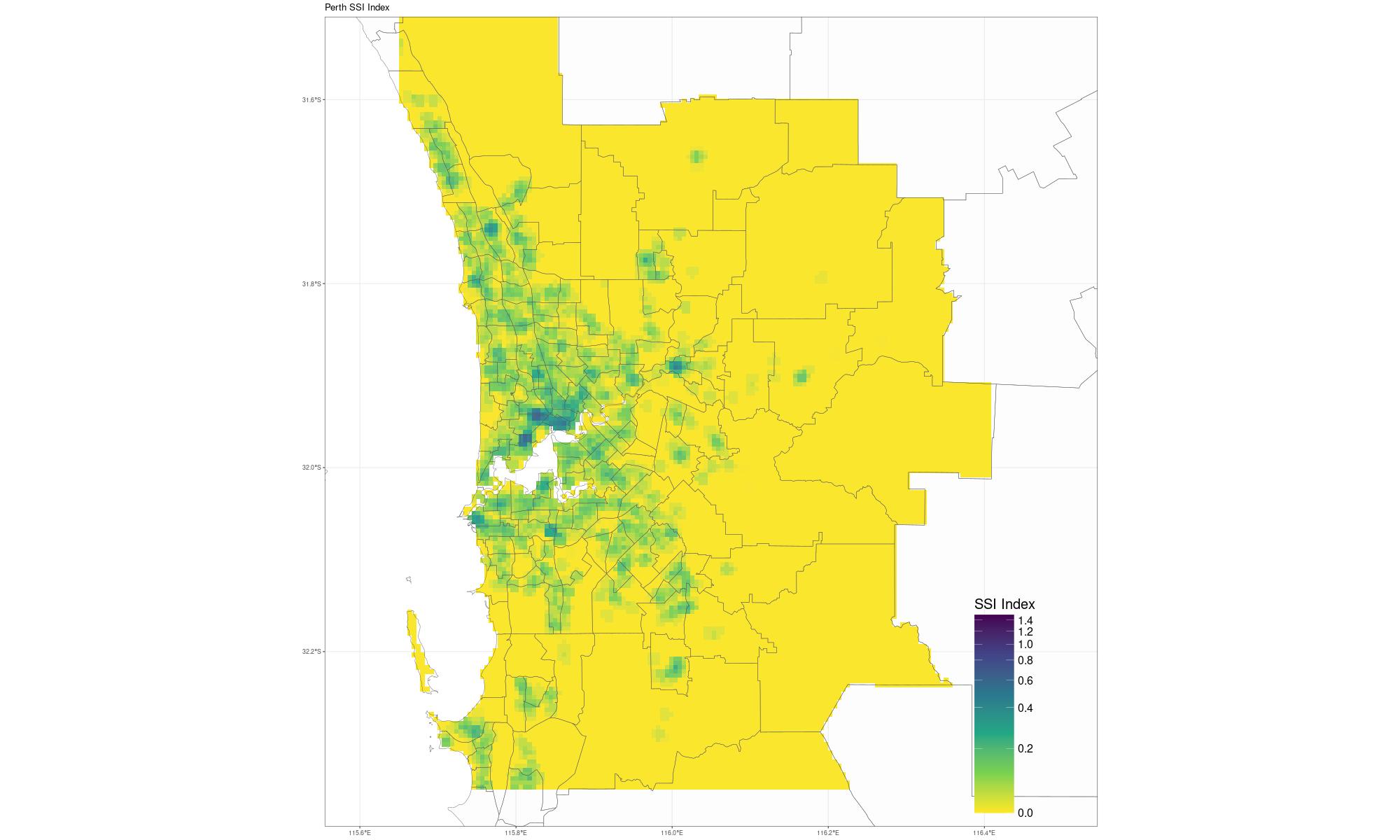"} \caption{Perth SSI}
\end{subfigure} %
\hfill
\begin{subfigure}[t]{0.3\textwidth}
\centering
\includegraphics[trim={260 0 260 20},clip,scale=0.11]{"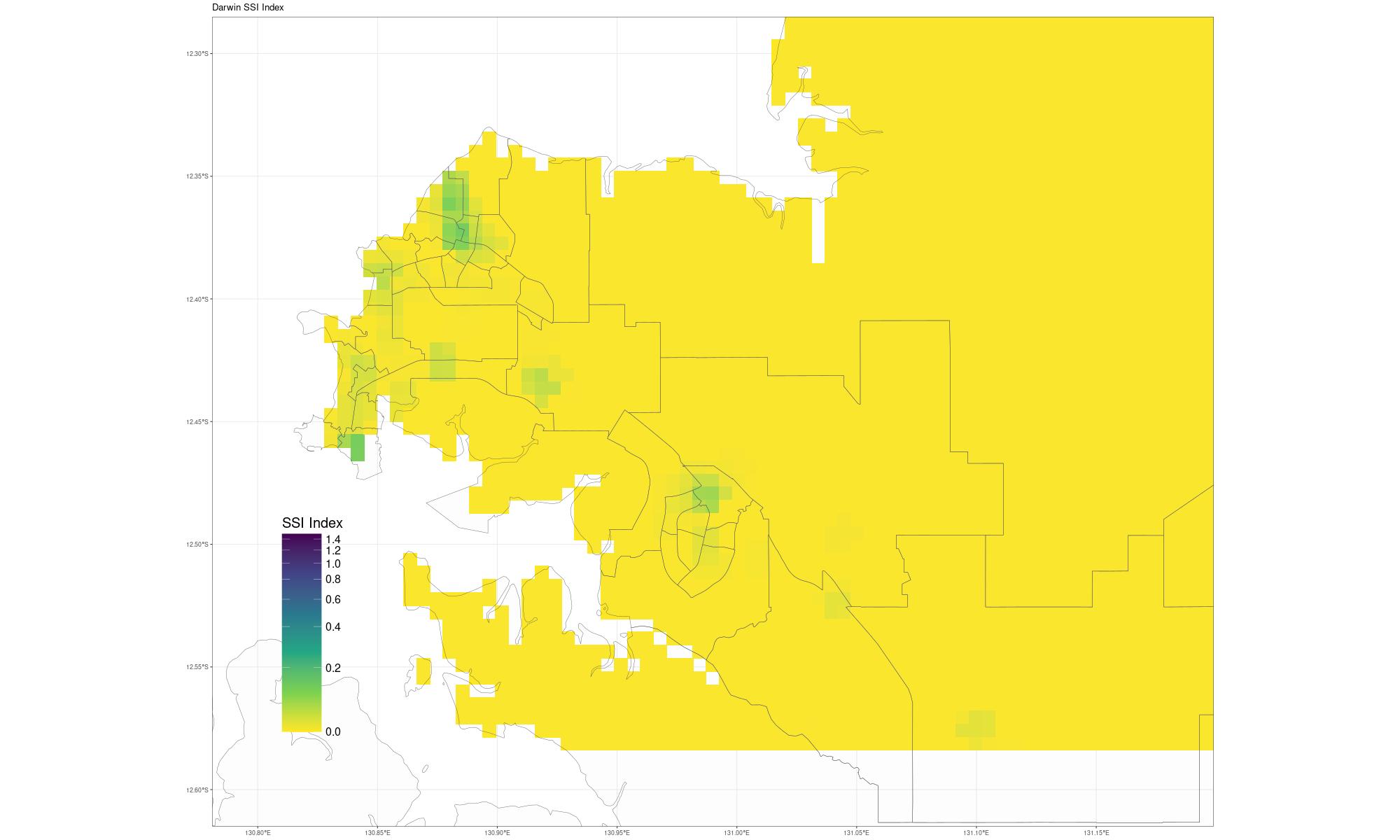"} \caption{Darwin SSI}
\end{subfigure}
\hfill
\begin{subfigure}[t]{0.3\textwidth}
\centering
\includegraphics[trim={630 0 630 20},clip,scale=0.11]{"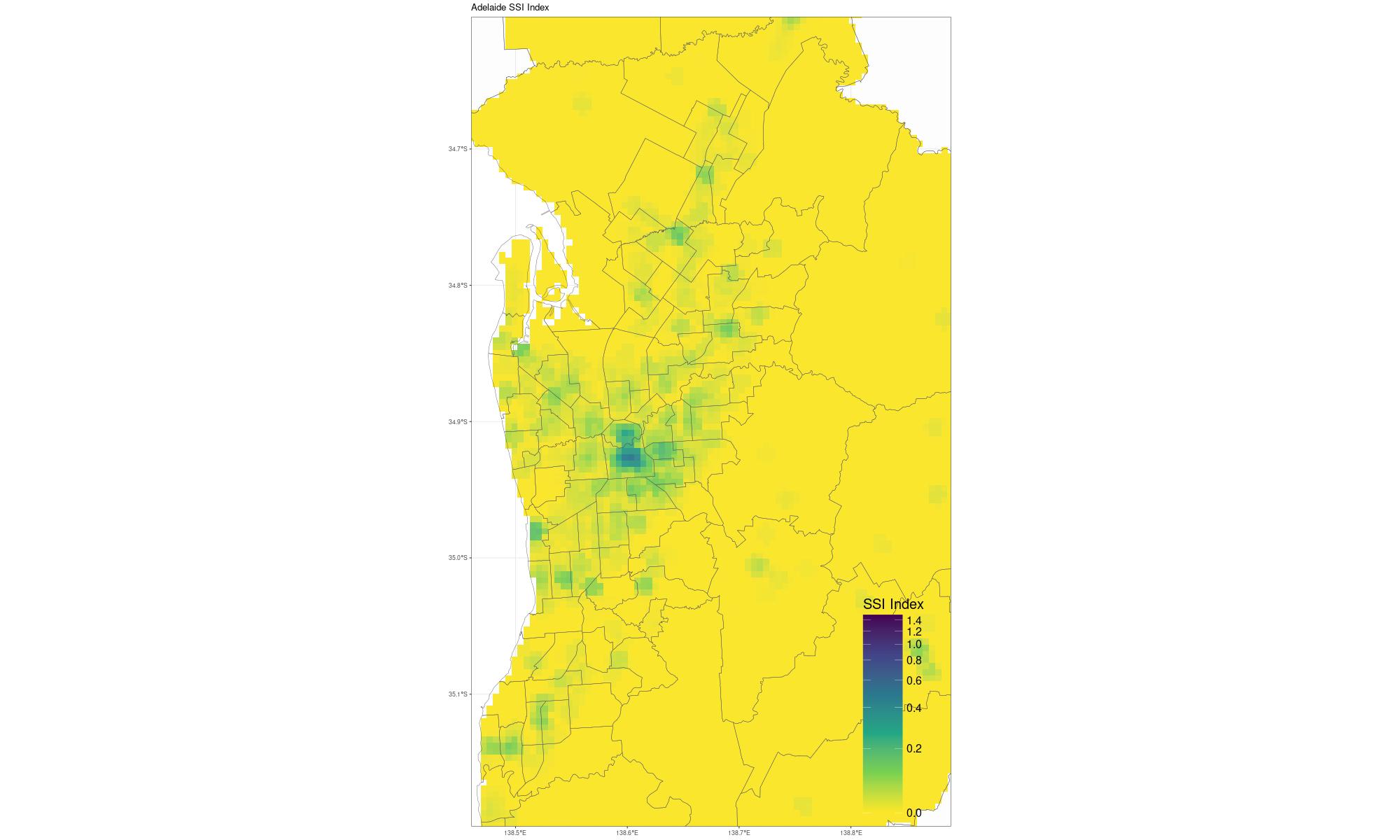"}\caption{Adelaide SSI}
\end{subfigure}
\caption{Calculated Social Service access Indexes, SSPT (including primary health care, childcare, and public transport frequency and proximity) and SSI (health care and childcare) at 100m mesh level for Melbourne, Sydney, Brisbane, Perth, Darwin, and Adelaide. Darker colors (blues/greens) represent higher values of access. Scales are logarithmic.}\label{fig:SSPTSSI}
\end{figure*}

\begin{figure*}[ht]
\centering
\begin{subfigure}[t]{0.49\textwidth}
\includegraphics[trim={428 0 200 0},clip,scale=0.16]{"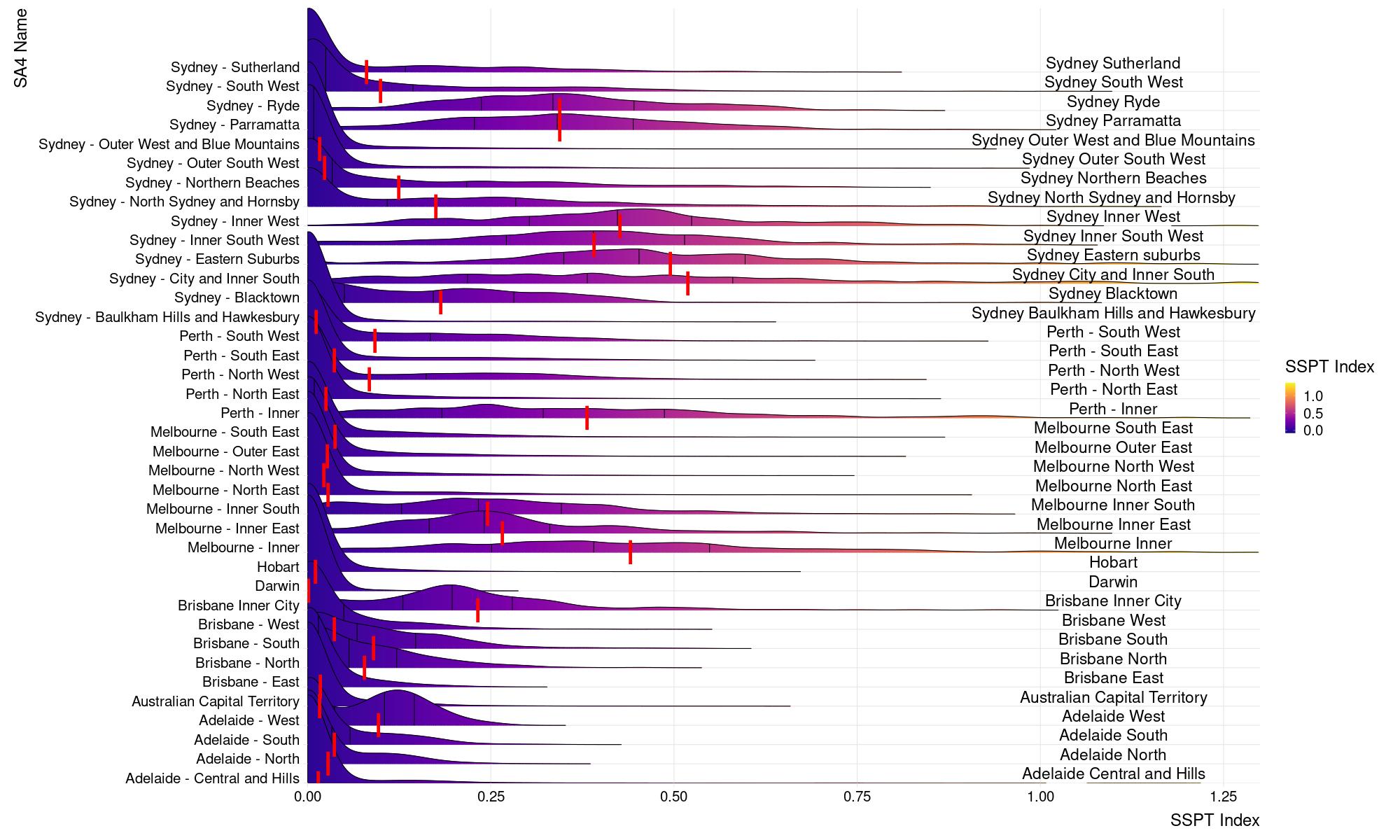"} \caption{SSPT index for SA4s}
\end{subfigure} %
\hfill
\begin{subfigure}[t]{0.49\textwidth}
\centering
\includegraphics[trim={428 0 180 0},clip,scale=0.16]{"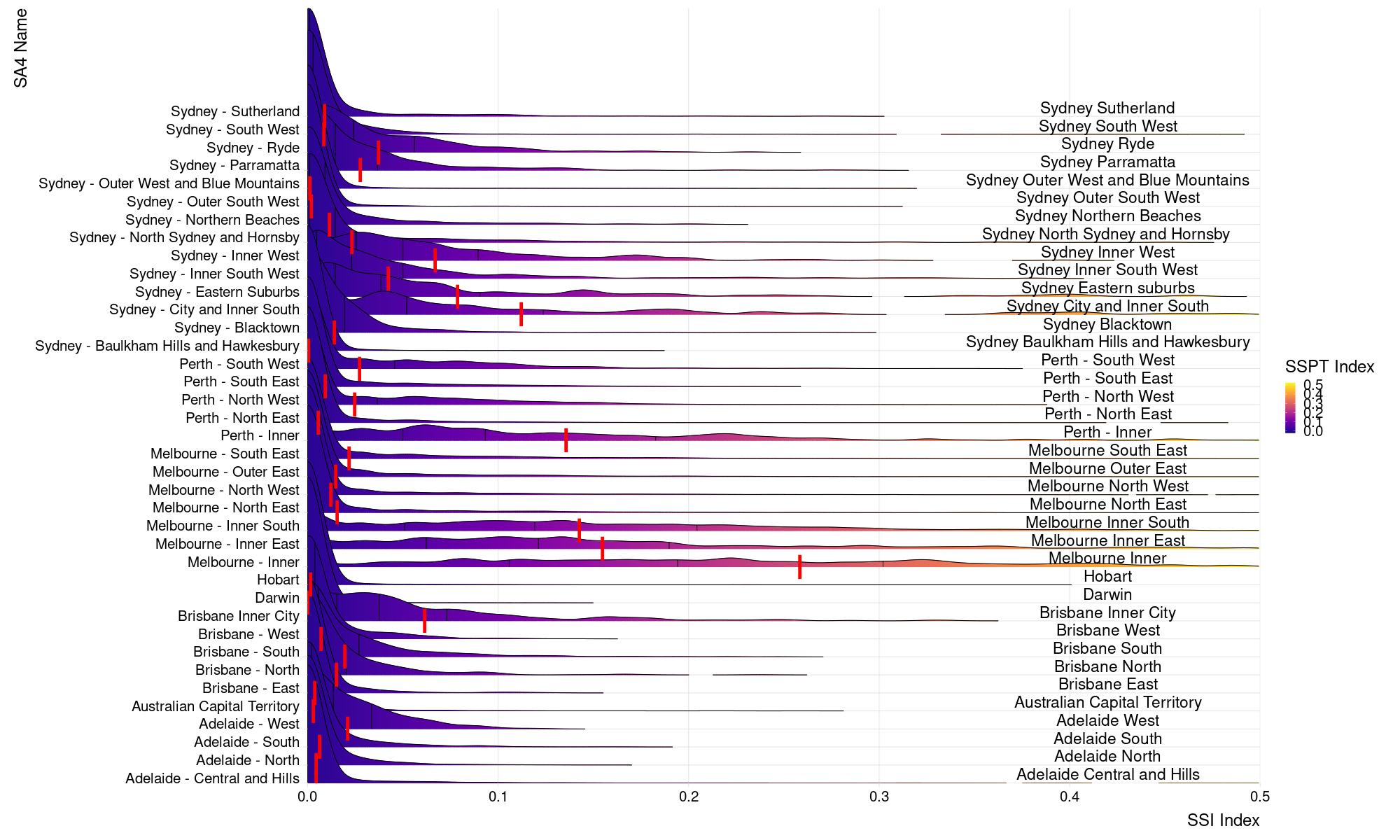"}  \caption{SSI index for SA4s}
\end{subfigure}
\hfill
\begin{subfigure}[t]{0.49\textwidth}
\centering
\includegraphics[trim={428 0 210 0},clip,scale=0.16]{"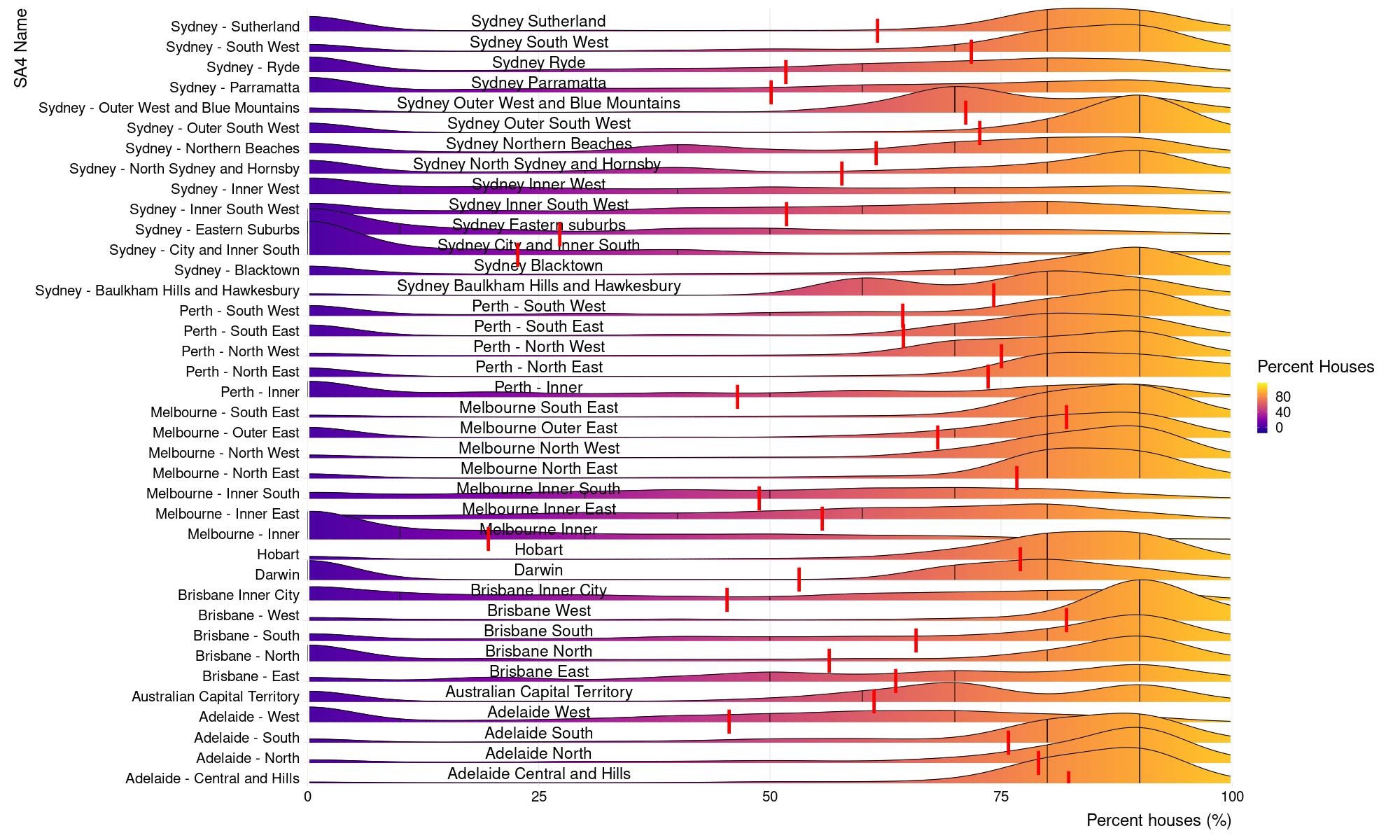"} \caption{Percent houses for SA4s}
\end{subfigure}
\centering
\begin{subfigure}[t]{0.49\textwidth}
 \includegraphics[trim={428 0 170 0},clip,scale=0.16]{"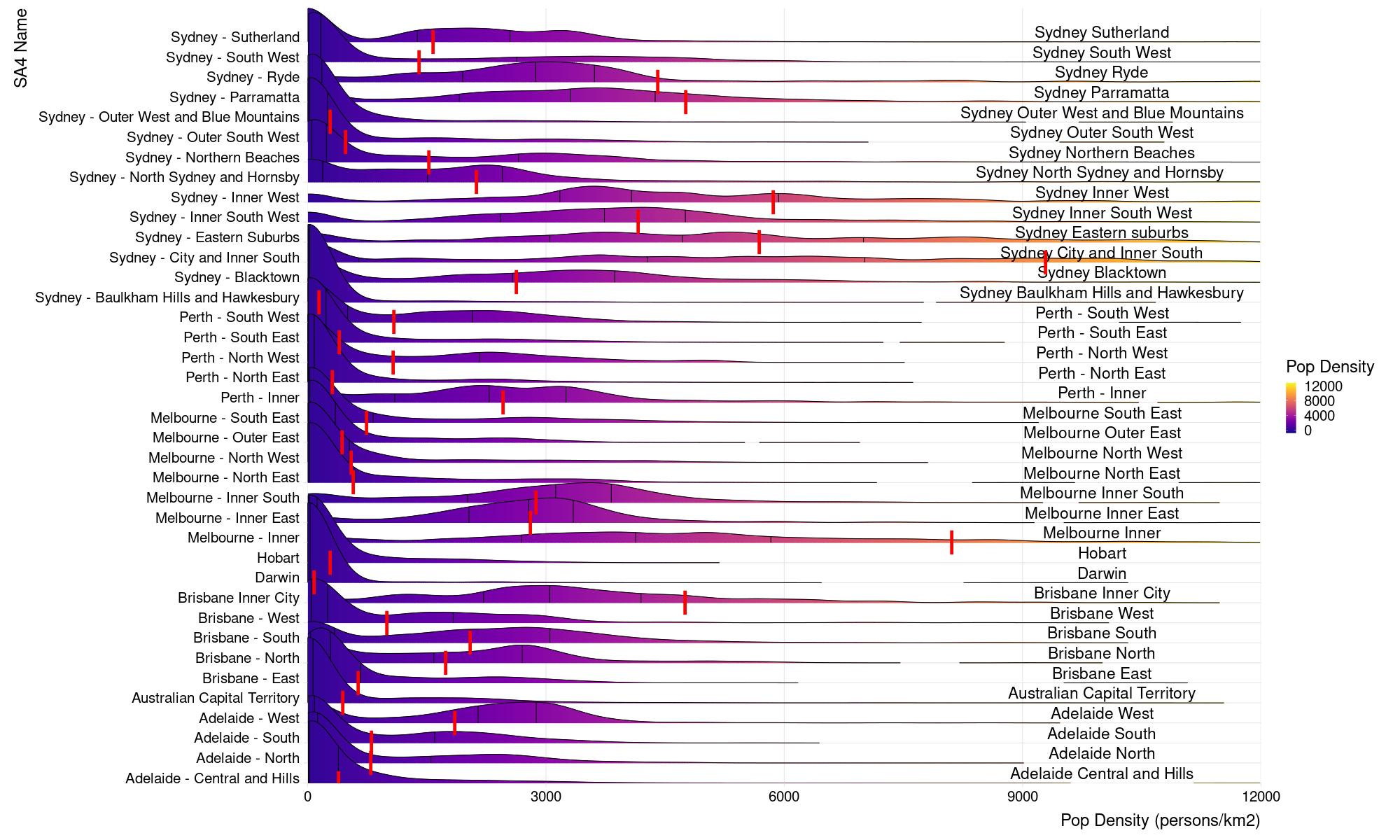"}\caption{Population density (persons/km2) for SA4s}
\end{subfigure} 
\caption{Distribution of 100m gridded SSPT and SSI index values and percentages of detached houses and population density across Sydney, Melbourne, Hobart, Darwin, Brisbane, Australian Capital Territory, and Adelaide SA4s. Mean values for each SA4 annotated with red tick.}\label{fig:sa4}
\end{figure*}

Maps of the estimated SSPT and SSI for six Australian capital cities are shown in Figure \ref{fig:SSPTSSI}. Figure \ref{fig:sa4} shows distributions of these index values across SA4 regions in these cities as well as population density and detached housing type percentages. They highlight the spatial and inequitable distributions of access to essential services such as public transport, primary health care and early childhood education that would be delivered by a compact city.

Access to social services and public transport in central urban areas, as measured by the SSPT index, are approximately 0.4-0.5 for Sydney, 0.25-0.4 for Melbourne, 0.25 for Brisbane, 0.15 for Perth and 0.10 Adelaide. In outer metropolitan areas of these cities and in the other cities (Darwin, Canberra, and Hobart), these scores quickly fall to 0.10 and below, indicating limited access to services and public transport. When only considering social services and excluding public transport access for inner-city areas, as measured by the SSI index, are approximately 0.15-0.25 for Melbourne, 0.15 for Perth, 0.10 for Sydney, 0.02 for Adelaide, and 0.01 for Brisbane. Similar to the SSPT index, even lower scores are seen in the outer areas of these cities and in the remaining Australian urban areas. 

Sydney, which is very well served by public transport proximity and service frequency, and that performs well in the SSPT index, shows greatly diminished access, as indicated by the SSI index, compared to Melbourne and Adelaide. This is likely due to a more fragmented topography compared to cities that have more contiguous geography. When further examining the individual components of the indexes, we find that access to childcare across the urban areas tends to be at a lower numbers than access to primary health care but more evenly distributed than primary health care services which in turn show higher overall numbers of total services and strong hotspot clustering, especially among transport corridors. This demonstrates that a single combined index can become overly influenced by a single factor and that the choice of factors and weightings should be carefully designed and determined by the intended research question to answer.

\begin{figure}[ht]
{\footnotesize a)}
\includegraphics[trim={520 0 520 20},clip,scale=0.14]{"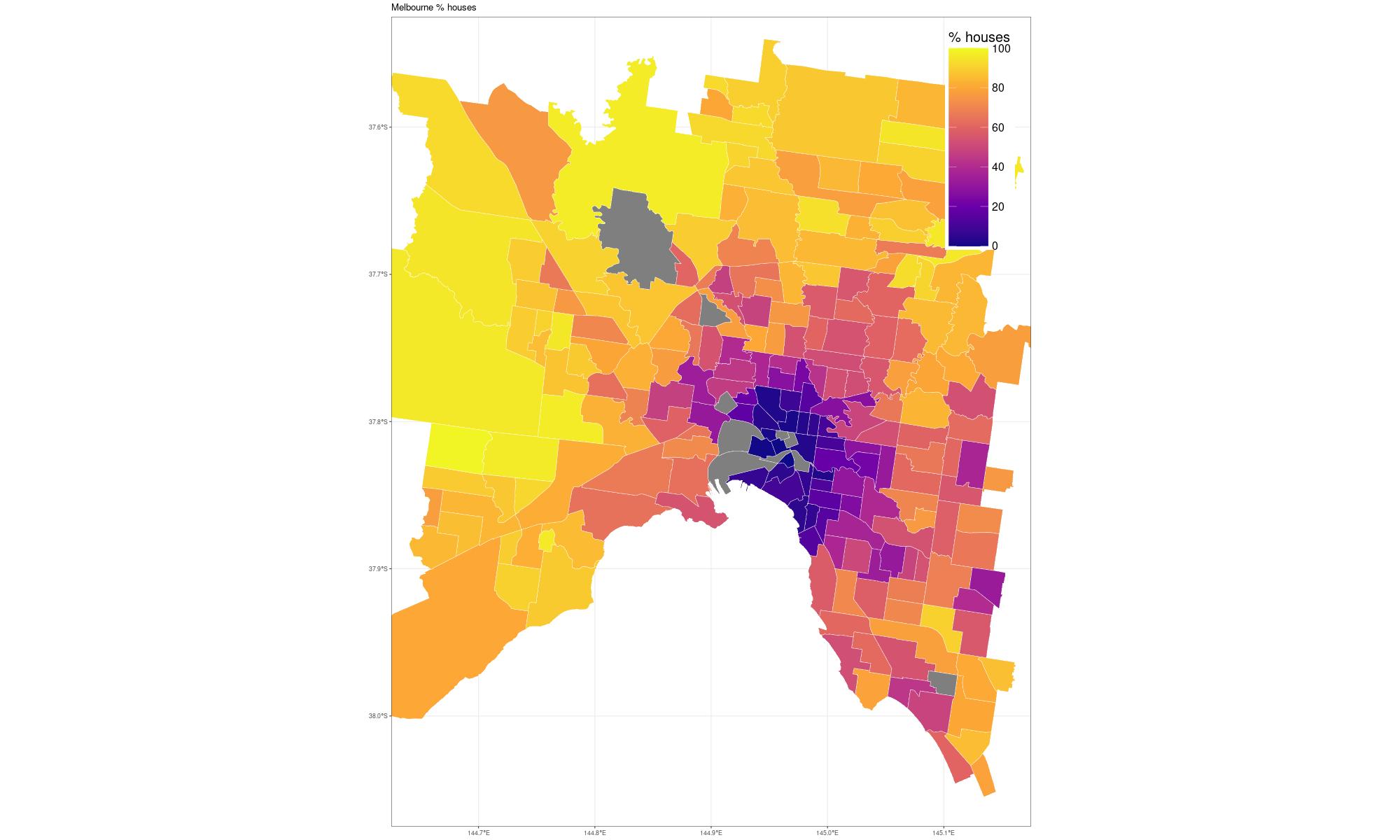"} {\footnotesize b)}\includegraphics[trim={520 0 520 20},clip,scale=0.14]{"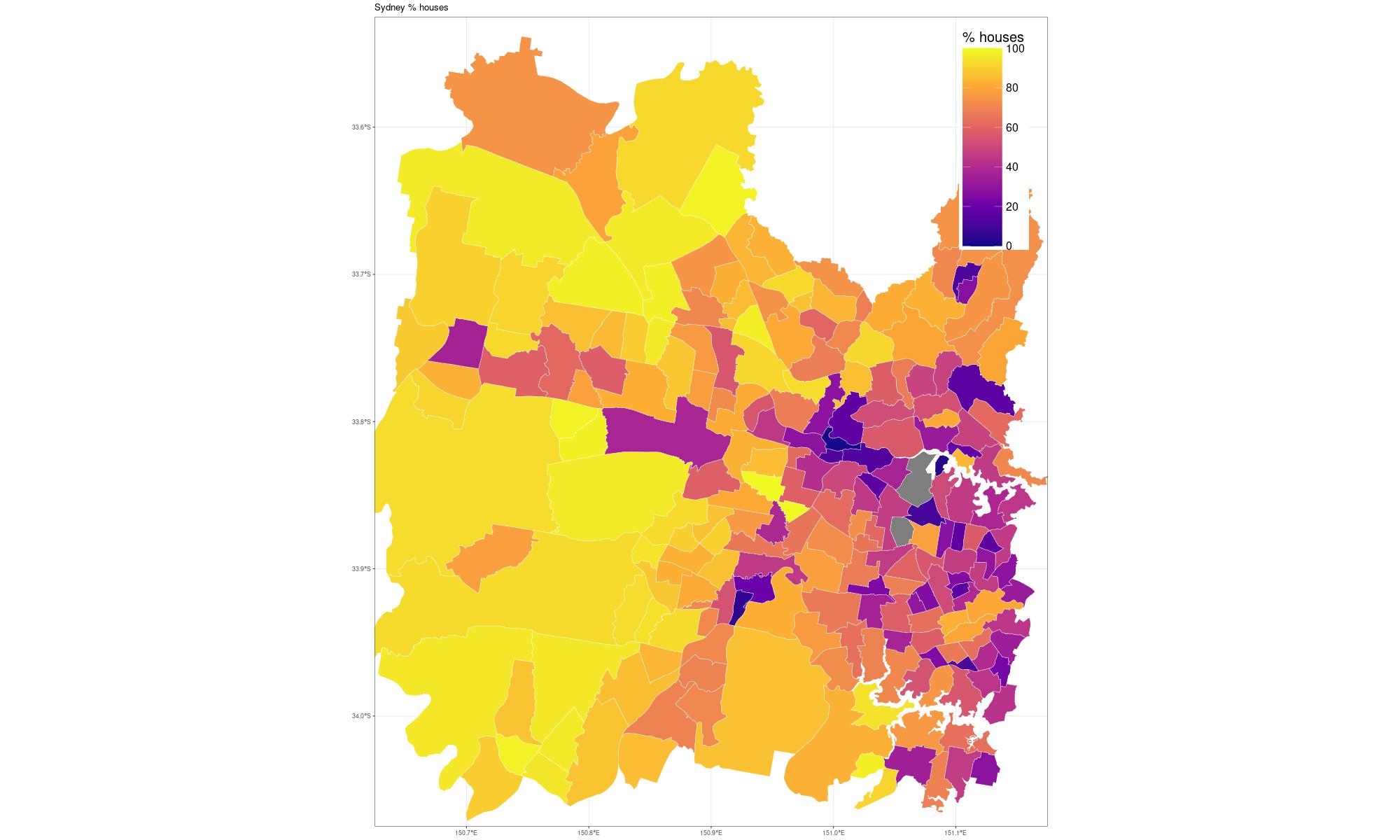"} 
{\footnotesize c)}\includegraphics[trim={520 0 485 20},clip,scale=0.14]{"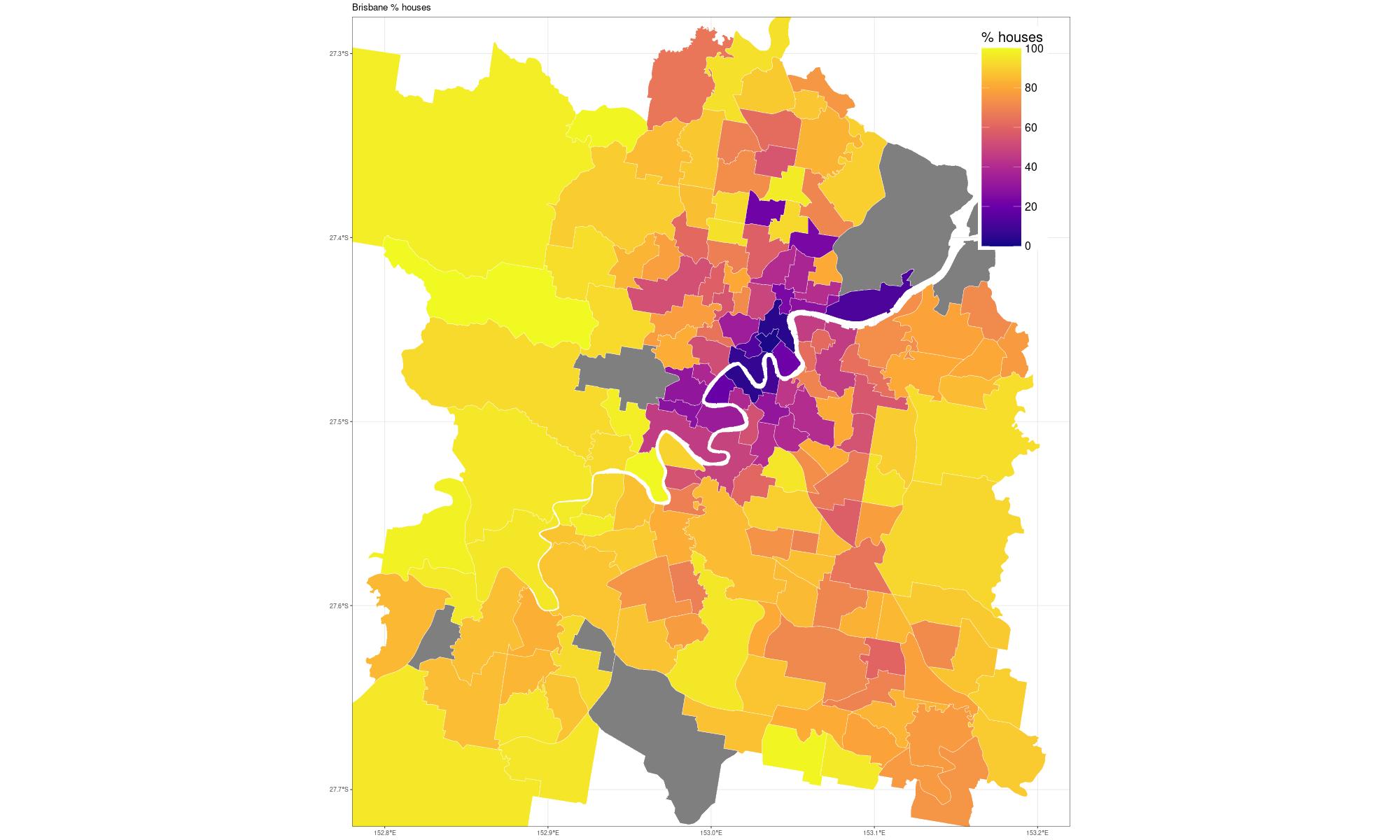"} 
{\footnotesize d)}\includegraphics[trim={520 0 505 20},clip,scale=0.14]{"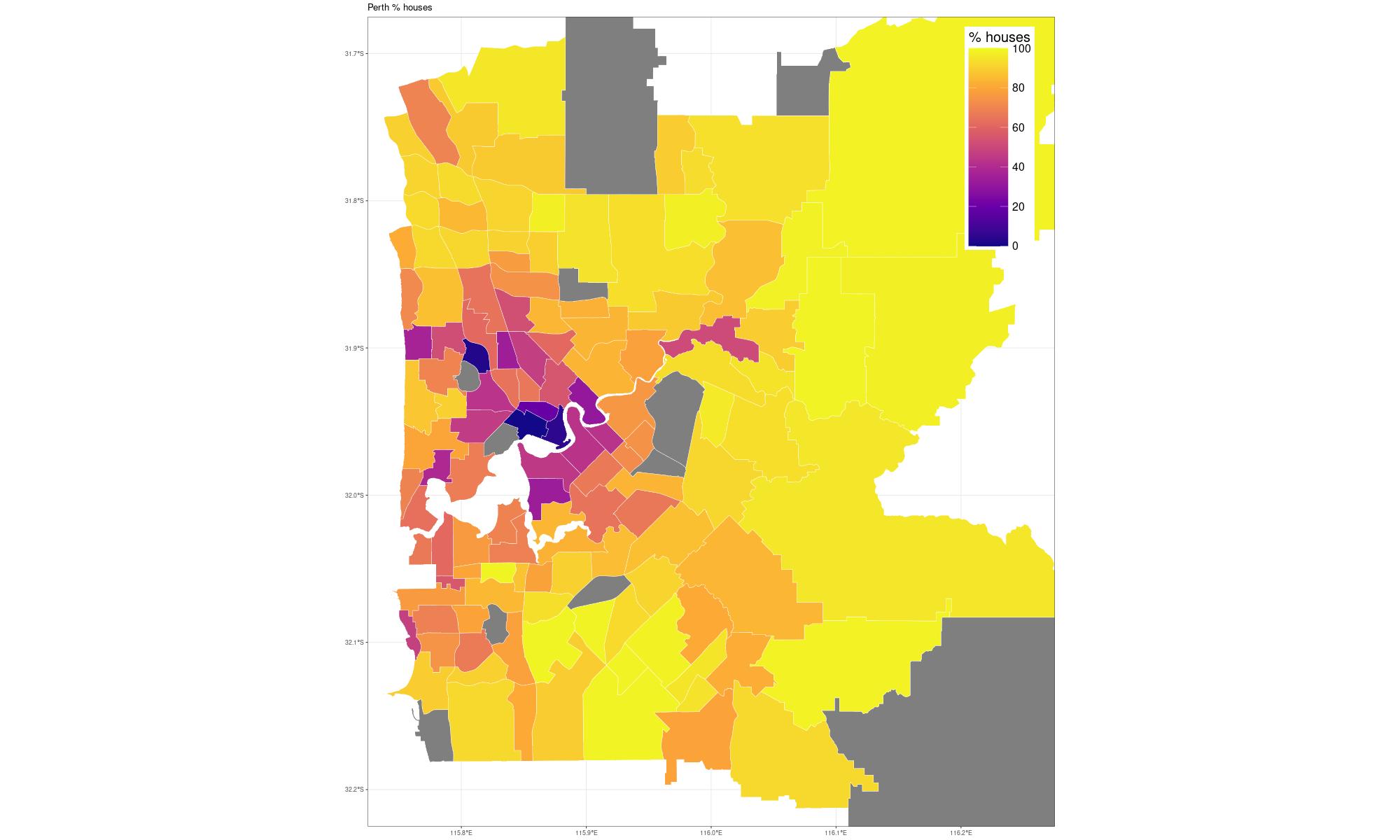"} 
{\footnotesize e)}\includegraphics[trim={520 0 443 20},clip,scale=0.14]{"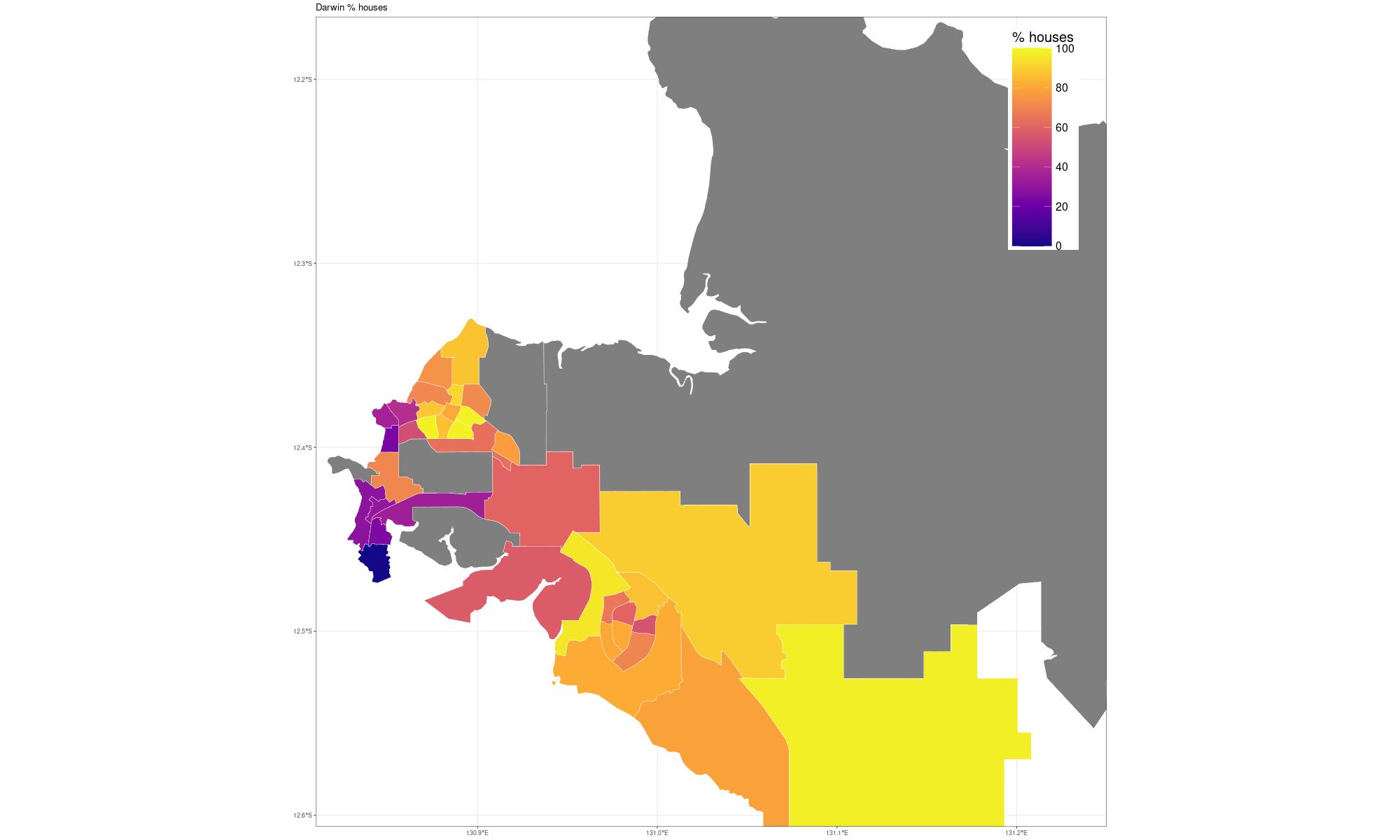"}  
{\footnotesize f)}\includegraphics[trim={520 0 522 20},clip,scale=0.14]{"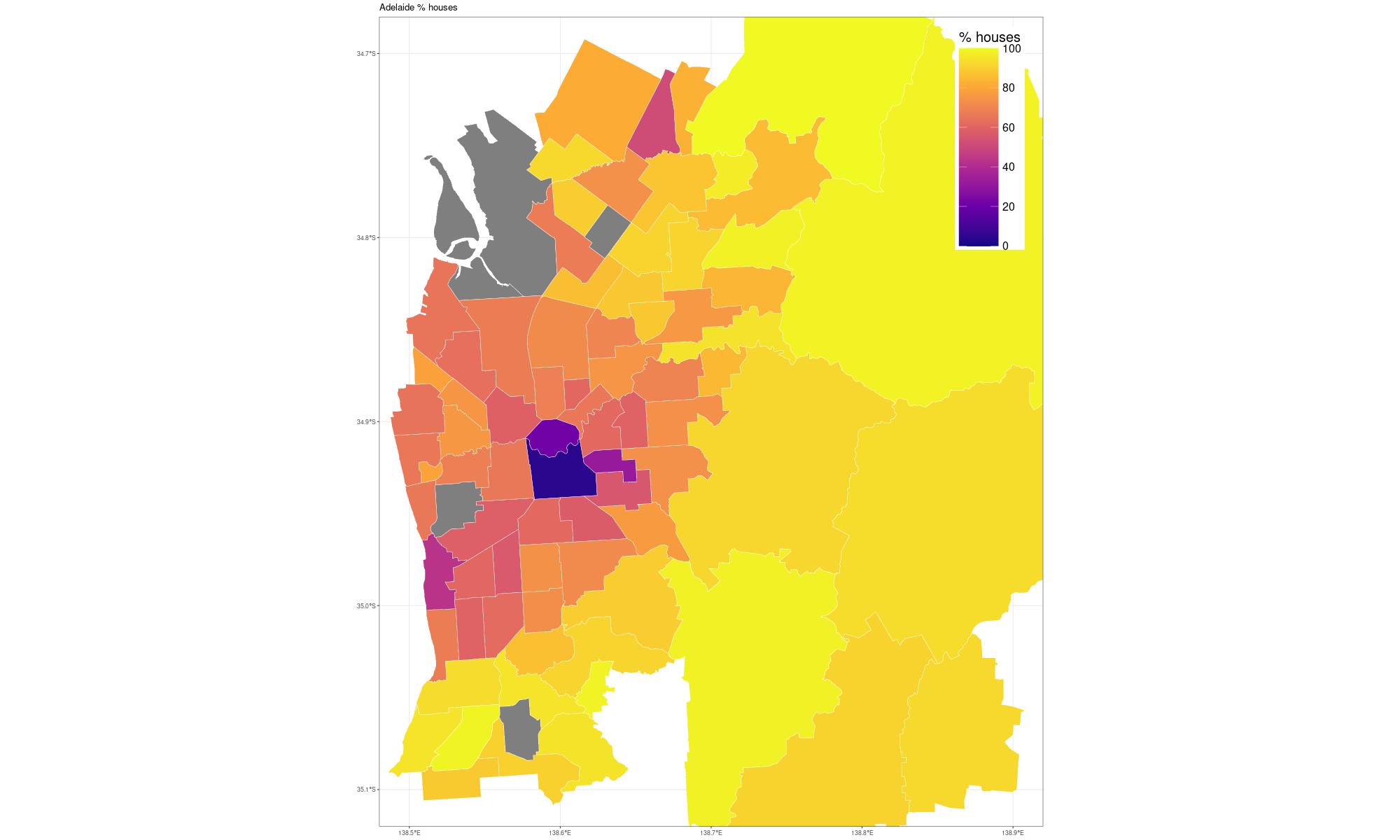"}
\caption{Housing types, showing percentages of detached houses for each SA2 for a) Melbourne, b) Sydney, c) Brisbane, d) Perth, e) Darwin, and f) Adelaide. Lighter colors indicate higher percentages of detached houses. Grey for areas with no housing/no data.}
\label{fig:houses}
\end{figure}

The proportion of property types (Figure \ref{fig:houses} and Table \ref{tab:sa4}) in each capital city was used as an additional proxy for urban density, percentages of detached houses compared to apartments. Other than Adelaide, Hobart and Darwin, the CBD area of Australia’s capital cities, comprise approximately 70\% of apartments as the main housing stock, reflecting the increased density in the monocentric Australian cities. Although access to services is delivered well in the densely housed and populated city centres, Australia's continued focus on monocentric planning \citep{gleeson_reviving_2015} delivers poor accessibility to essential services. Access to social services is especially limited particularly in Adelaide, Perth, Brisbane, Hobart, Canberra and Darwin where housing is predominately single detached dwellings.

%

At a broad level, Table \ref{tab:sa4} and Figures \ref{fig:SSPTSSI} and \ref{fig:sa4} shows that urban regions (SA4s) with the highest accessibility are generally those with the greatest population density and the lowest percentage of standalone housing and highest percentage of apartments. These areas generally correspond to the inner city regions. In Melbourne, this consists of a single inner city region, where Sydney has a number of dense (but lower density) inner regions spread across the city. None of the other Australian cities have levels of density and access approaching Melbourne and Sydney.

\begin{table}[ht!]
		\caption{\bf{Social service index results aggregated to SA4 levels in Australian capital cities. Other demographic information including population density (persons/km$^{2}$), housing types (percent), and socioeconomic indexes (SEIFA) have also been aggregated to a SA4 level.}}  
		\label{tab:sa4} 
		\centering
		\begin{tabular}{|p{1.41cm}|p{3.70cm}|p{0.8cm}|p{0.9cm}|p{0.9cm}|p{0.9cm}|p{1.2cm}|p{0.65cm}|p{0.9cm}|}
		\hline
		City & SA4 & SSPT Index & SSI Index & Density & Houses & Semi detached & Apt. & SEIFA \\ \hline		
 	    Adelaide &  West & 0.10 & 0.021 & 1857 & 66&22&11 & 43\\ \hline
		 &  South & 0.04 & 0.007 & 810 & 79&15&6 & 65\\ \hline
		 &  North & 0.03 & 0.005 & 804 & 83&13&3 & 46\\ \hline
		 &  Central and Hills & 0.02 & 0.005 & 393 & 66&18&15 & 72\\ \hline		 
		 
		 Canberra & Australian Capital Territory & 0.017 & 0.003 & 448 & 84&9&6 & 68\\ \hline	
		 	 
		 Brisbane &  Inner City & 0.24 & 0.062 & 4758 & 35&7&58 & 80\\ \hline
		  &  South & 0.09 & 0.02 & 2051 & 69&14&15 & 61\\ \hline
		  &  North & 0.08 & 0.016 & 1741 & 69&14&17 & 64\\ \hline
		  &  West & 0.04 & 0.007 & 1001 & 77&7&17 & 84\\ \hline 		 
		  &  East & 0.02 & 0.004 & 640 & 82&12&5 & 43\\ \hline
		  		 
		 Darwin & Darwin & 0.002 & 0.0005 & 86 & 63&10&25 & 46\\ \hline	
		 	 
		 Hobart & Hobart & 0.011 & 0.002 & 294 & 85&6&8 & 54\\ \hline	
		 	 
		 Melbourne &  Inner & 0.44 & 0.26 & 8120 & 20&22&57 & 73\\ \hline
		  &  Inner South & 0.25 & 0.143 & 2885 & 53&24&22 & 75\\ \hline
		  &  Inner East & 0.27 & 0.155 & 2810 & 60&21&19 & 73\\ \hline
		  &  South East & 0.038 & 0.022 & 745 & 81&12&6 & 58\\ \hline
		  &  North East & 0.028 & 0.016 & 585 & 80&15&5 & 65\\ \hline
		  &  North West & 0.022 & 0.012 & 553 & 81&16&3 & 65\\ \hline
		  &  Outer East & 0.027 & 0.015 & 442 & 86&11&2 & 64\\ \hline
		  		
		Perth &  Inner & 0.38 & 0.14 & 2472 & 48&15&36 & 81\\ \hline
		  &  South West & 0.093 & 0.027 & 1094 & 82&12&6 & 56\\ \hline
		  &  North West & 0.084 & 0.025 & 1088 & 78&18&4 & 59\\ \hline
		  &  South East & 0.037 & 0.010 & 408 & 80&14&6 & 55\\ \hline
		  &  North East & 0.025 & 0.006 & 319 & 84&12&4 & 62\\ \hline
		  		
		Sydney &  City and Inner South & 0.52 & 0.11 & 9299 & 10&19&69 & 60\\ \hline
		 &  Inner West & 0.43 & 0.067 & 5866 & 34&16&49 & 65\\ \hline	 
		 &  Eastern Suburbs & 0.49 & 0.079 & 5694 & 23&18&58 & 80\\ \hline
		 &  Parramatta & 0.34 & 0.028 & 4767 & 47&14&39 & 42\\ \hline
		 &  Ryde & 0.34 & 0.037 & 4412 & 47&13&39 & 75\\ \hline
		 &  Inner South West & 0.39 & 0.043 & 4172 & 51&14&34 & 39\\ \hline
		 &  Blacktown & 0.18 & 0.014 & 2635 & 81&12&7 & 48\\ \hline	
		 &  North Sydney and Hornsby & 0.17 & 0.023 & 2136 & 46&8&46 & 86\\ \hline
		 &  Sutherland & 0.081 & 0.009 & 1588 & 61&14&25 & 80\\ \hline		 	
   	     &  Northern Beaches & 0.13 & 0.012 & 1531 & 57&9&33 & 86\\ \hline	
		 &  South West & 0.10 & 0.009 & 1409 & 75&11&14 & 44\\ \hline
		 &  Outer South West & 0.02 & 0.002 & 489 & 84&11&4 & 60\\ \hline
		 &  Outer West and Blue Mountains & 0.017 & 0.001 & 289 & 82&11&7 & 64\\ \hline
		 &  Baulkham Hills and Hawkesbury & 0.012 & 0.0007 & 147 & 83&8&8 & 60\\ \hline

		\hline  
		    \end{tabular}
		\end{table}

In answering our questions, we show that two Australian cities (Melbourne and Sydney) have some characteristics of a compact or 15-minute city, generally in the CBD and the inner suburbs where greater housing density is situated. However, even in them, access quickly diminishes in the middle and outer regions. If we intend to start defining what levels of access are considered adequate, as \cite{ryan_defining_2023} urge is necessary and that is something policy makers are often reluctant to do, even minimal goals such as at least one childcare facility, one pharmacy, and one GP within a 800m distance of all urban residents are currently beyond the reach of most in Australian cities. Overall, access to social services is limited and bereft in the peri-urban areas of Australian cities, areas with lower socioeconomic demographics, thereby exacerbating social inequalities. 

These results are in line with \cite{bruno_universal_2024}'s global assessment of 15-minute cities which found that no Australian cities can be considered 15-minute cities. Melbourne and Sydney performed best, with walking trips to services estimated at 17 minutes and 19 minutes respectively. This suggests that access to essential services such as health care and mental health support is suboptimal and this is particularly the case for populations without access to private transport. 

While our findings provide a baseline analysis of the range of spatial advantages and disadvantages experienced in accessing some essential services across Australian cities, future research should also expand to include other services such as groceries, community hubs, and employment and explore not only the duration and lengths of journeys to these services but also assess their quality.

\section{Acknowledgements}

This research was supported by the Australian Research Council through the Centre of Excellence for Children and Families over the Life Course (Life Course Centre), https://lifecoursecentre.org.au/

\section{Data Availability}
Data is available at: https://doi.org/10.5281/zenodo.17587608
\\
Code is available at: https://gitlab.unimelb.edu.au/knice/urbanaccessindex/

\bibliographystyle{apa}
\bibliography{bibliography.bib}

\end{document}